\documentclass[12pt,preprint]{elsarticle}
\usepackage{graphicx}
\usepackage{adjustbox}
\usepackage{amsmath,amssymb}
\usepackage{lineno}
\journal{Phys.\,Lett.\,B (17 Dec. 2019)}

\usepackage{rotating}
\usepackage{texnames}
\usepackage[T1]{fontenc}
\usepackage{hepnames}
\usepackage{siunitx}
\usepackage{color}
\usepackage{nicefrac}
\usepackage{hyperref}
\hypersetup{colorlinks = true, allcolors = blue}
\usepackage{bold-extra}
\usepackage{soul}
\usepackage{multirow}

\begin{document}

\begin{frontmatter}

\title{Improved Bhabha cross section at LEP\\ and the number of light neutrino species\\
{\footnotesize \it Marking the 30th anniversary of the first \texorpdfstring{$N_\nu$}{nnu} determination at LEP on 13 October 1989}}

\author[1]{Patrick Janot}
\ead{Patrick.Janot@cern.ch}
\author[2]{Stanis\l{}aw Jadach}
\ead{Stanislaw.Jadach@cern.ch}

\address[1]{CERN, EP Department, 1 Esplanade des Particules, CH-1217 Meyrin, Switzerland} 
\address[2]{Institute of Nuclear Physics PAN, ul. Radzikowskiego 152, 31-342 Krak\'ow, Poland}


\begin{abstract}

{\small 
In $\rm e^+ \rm e^-$ collisions, the integrated luminosity is generally measured from the rate of low-angle  Bhabha interactions $\rm e^+ e^- \to e^+ e^- $. In the published LEP results, the inferred theoretical uncertainty of $\pm 0.061\%$ on the predicted rate is significantly larger than the reported experimental uncertainties. We present an updated and more accurate prediction of the Bhabha cross section in this letter, which is found to reduce the Bhabha cross section by about $0.048\%$, and its uncertainty to $\pm 0.037\%$. When accounted for, these changes modify the number of light neutrino species (and its accuracy), as determined from the LEP measurement of the hadronic cross section at the Z peak, to $N_\nu = 2.9963 \pm 0.0074$. The 20-years-old $2 \sigma$ tension with the Standard Model is gone.} 

\end{abstract}


\date{Revised 13 Feb. 2020}

\end{frontmatter} 

\vfill\eject

\section{Introduction}
\label{sec:introduction}

The Large Electron-Positron (LEP) collider was operated at CERN between $1989$ and $2000$, and delivered $\rm e^+ e^-$ collisions to four experiments, at centre-of-mass energies that  covered the $\rm Z$ resonance, the $\rm WW$ threshold, and extended up to $\sqrt{s} = 209$\,GeV. The first phase (LEP1), at and around the $\rm Z$ pole, provided a wealth of  measurements of unprecedented accuracy~\cite{ALEPH:2005ab}. In particular, the measurement of the hadronic cross section at the Z peak, $\sigma_{\rm had}^0$, has been used to derive the number of light neutrino species $N_{\nu}$ from
\begin{equation}
N_\nu \left( \frac{\Gamma_{\nu\nu}}{\Gamma_{\ell\ell}}\right)_{\rm SM}= \left( \frac{12\pi}{m_{\rm Z}^2} \frac{R^0_\ell}{\sigma^0_{\rm had}} \right)^{\frac{1}{2}} - R^0_\ell - 3 - \delta_\tau, 
\label{eq:NnuForLep}
\end{equation}
where $R^0_\ell$ is the ratio of the hadronic-to-leptonic Z branching fractions; the correction $\delta_\tau \simeq -2.263 \times 10^{-3}$~\cite{DUBOVYK201886} accounts for the effect of the $\tau$ mass on the ${\rm Z} \to \tau\tau$ partial width; and $(\Gamma_{\nu\nu}/\Gamma_{\ell\ell})_{\rm SM}$ is the ratio of the massless neutral-to-charged leptonic Z partial widths predicted by the Standard Model (SM). The combination of the measurements made by the four LEP experiments~\cite{ALEPH:2005ab} led to:
\begin{equation}
   N_{\nu} = 2.9840 \pm 0.0082,
   \label{eq:Nnu}
\end{equation}
consistent within two standard deviations with the three observed families of fundamental fermions.\footnote{In Ref.~~\cite{ALEPH:2005ab}, the right part of Eq.~\ref{eq:NnuForLep} was evaluated to be $5.943 \pm 0.016$ from a combination of the LEP measurements, and  $(\Gamma_{\nu\nu}/\Gamma_{\ell\ell})_{\rm SM}$ was determined in the SM to amount to $1.99125 \pm 0.00083$, leading actually to $N_\nu = 2.9846 \pm 0.0082$. The difference of $0.0006$ with Ref.~\cite{ALEPH:2005ab} is attributed to rounding effects~\cite{VerteForet}.
}
This observable is directly affected by any systematic bias on the integrated luminosity through $\sigma^0_{\rm had}$. Indeed, the integrated luminosity uncertainty dominates the uncertainty on $\sigma^0_{\rm had}$, and is the largest contribution to the $N_{\nu}$ uncertainty. For example, collective beam-induced effects were recently discovered to produce a bias of almost $-0.1\%$ on the LEP integrated luminosity, 
and thus to increase the number of light neutrino species by 
$\delta N_\nu = +0.0072 \pm 0.0004$~\cite{Voutsinas:2019hwu}.

At LEP, the luminosity was determined by measuring the rate of the theoretically well-understood Bhabha-scattering process at small angles, $\rm e^+ e^- \to \rm e^+ \rm e^- $, in a set of dedicated calorimeters (LumiCal), possibly completed with tracking devices,  situated close to the beam axis on each side of the interaction region. The Bhabha events were selected with a ``narrow'' acceptance on one side and a ``wide'' acceptance on the other, defined as shown in Table~\ref{tab:AngularRanges} for the LumiCals used during the LEP\,1 phase between 1990 and 1995. Such an asymmetric acceptance strongly damps the sensitivity to effects that tend to modify the measured acollinearity distribution of the outgoing electrons and positrons: photon emission from initial-state radiation or beamstrahlung, position or energy spread at the interaction point, misalignments, etc. 
\begin{table}[htbp]
\small\centering
\vspace{-3mm}
\caption{Wide and narrow acceptance for first- and second-generation LumiCals of the four LEP experiments. The periods where these devices were operated are also indicated.  The ALEPH LCAL numbers are only indicative, as the fiducial acceptance followed the (square) detector cell boundaries, instead of specific polar angle values. The detector emulation used in this paper includes this subtlety.%
\vspace{1mm} }
\label{tab:AngularRanges}
\begin{tabular}{|l|c|c|c|c|}
    \hline
 \multirow{ 2}{*}{Expt / LumiCal} & \multirow{ 2}{*}{Period} & Narrow & Wide \\
                           &                          & (mrad) & (mrad)          \\ \hline\hline
  ALEPH LCAL~\cite{Decamp:224392}             & 01/90 $\to$ 08/92 & 57\phantom{.0} -- 107\phantom{.0} & 43 -- 125\\ \hline 
  DELPHI SAT~\cite{Bugge:2629111,AlexPrivate} & 01/90 $\to$ 12/93 & 56.0 -- 128.6 & 52.7 -- 141.8 \\ \hline 
  L3 BGO~\cite{Acciarri:261072} & 01/90 $\to$ 12/92 & 31.2 -- \phantom{1}65.2 & 25.2 -- \phantom{1}71.2 \\ \hline
  OPAL FD~\cite{Acton:245031}  & 01/90 $\to$ 12/92 & 65.0 -- 105.0 & 55.0 -- 115.0 \\ \hline\hline  
  ALEPH SiCAL~\cite{Buskulic:1994wz} & 09/92 $\to$ 12/95 & 30.4 -- \phantom{1}49.5 & 26.1 -- \phantom{1}55.9 \\ \hline 
  DELPHI STIC~\cite{Alvsvaag:365298} & 01/94 $\to$ 12/95 & 43.6 -- 113.2 & 37.2 -- 126.8 \\ \hline 
  L3 SLUM~\cite{Brock:307585} & 01/93 $\to$ 12/95 & 32.0 -- \phantom{1}54.0 & 27.0 -- \phantom{1}65.0 \\ \hline
  OPAL SiW~\cite{Abbiendi:1999zx} & 01/93 $\to$ 12/95 & 31.3 -- \phantom{1}51.6 & 27.2 -- \phantom{1}55.7 \\ \hline
\end{tabular}
\end{table}

Any precision improvement in the prediction of the Bhabha cross section in the LumiCal acceptance would translate in a more accurate determination of the integrated luminosity, and therefore of the number of light neutrino species. The LEP experiments consistently evaluated the Bhabha cross section and their event selection efficiency with the Monte-Carlo event generator {\tt BHLUMI}~\cite{Jadach:225857,Jadach:1996is} in different versions (Table~\ref{tab:BHLUMIVersion}), with the following caveat. Until the end of 1992, ALEPH~\cite{Barate:1999ce}, DELPHI~\cite{Abreu:258291,Abreu:259634}, and L3~\cite{Acciarri:261072} corrected for the Z exchange contribution with event re-weighting obtained from the {\tt BABAMC} generator, while OPAL~\cite{Abbiendi:2000hu} scaled their 1990-92 luminosity values to the {\tt BHLUMI 4.04} predictions
in their last publication. The corresponding theoretical uncertainties on the quoted integrated luminosity, typically 0.20-0.30\% with {\tt BHLUMI 2.01} in the first-generation LumiCals, and 0.061\% with {\tt BHLUMI 4.04} in the second-generation LumiCals, are also indicated in Table~\ref{tab:BHLUMIVersion}. The uncertainty breakdown is displayed in Table~\ref{tab:breakdown}. 

\begin{table}[htbp]
\small
\centering
\caption{Versions of {\tt BHLUMI} used throughout the LEP\,1 phase. In 1990, ALEPH~\cite{Barate:1999ce} used the {\tt BABAMC} generator~\cite{Berends:1987jm} instead of {\tt BHLUMI}. The corresponding uncertainty on the Bhabha cross section, as quoted by each experiment, is indicated in brackets. %
\vspace{1mm}}
\label{tab:BHLUMIVersion}
\begin{tabular}{|r|c|c|c|c|}
    \hline
               &   ALEPH & DELPHI & L3 &  OPAL  \\ \hline\hline
     1990  & {\tt BABAMC} {\small (0.320\%)} & \multirow{3}{*}{{\tt 2.01} {\small (0.300\%)}} & \multirow{3}{*}{{\tt 2.01} {\small (0.250\%)}} & {\tt 2.01} {\small (0.300\%)} \\ \cline{1-2} 
    1991-92 & {\tt 2.01} {\small (0.210\%)} &  &  & {\small later scaled} \\ \cline{1-2}
    Fall 92 & {\tt 2.01} {\small (0.160\%)} &  & & to {\tt 4.04} \\ \hline
    1993 & \multirow{2}{*}{{\tt 4.04} {\small (0.061\%)}} & {\tt 4.02} {\small (0.170\%)} & \multirow{2}{*}{{\tt 4.04} {\small (0.061\%)}} & \multirow{2}{*}{{\tt 4.04} {\small (0.054\%)}}\\ \cline{1-1} \cline{3-3}
    1994-95 & & {\tt 4.03} {\small (0.061\%)} & & \\ \hline
\end{tabular}
\end{table}

The theoretical progress made between versions {\tt 2.01} and {\tt 4.04} of {\tt BHLUMI} include an improved 
implementation of the QED corrections to the Z exchange diagram contribution (improved with respect to {\tt BABAMC} as well)~\cite{Jadach:1995hv}; and a better estimate~\cite{Burkhardt:1995tt,Eidelman:1995ny} of the vacuum polarization in the $t$-channel photon propagator, instead of that of Ref.~\cite{Burkhardt:201068} in version {\tt 2.01}. In addition, the main QED matrix element of {\tt BHLUMI} {\tt 4.04} includes non-soft ${\cal O}(\alpha^2 L_{\rm e}^2)$ corrections, with $L_{\rm e}=\ln(|t|/m_{\rm e}^2)$ due to photons, absent from {\tt BHLUMI 2.01}, while {\tt BABAMC} implements pure ${\cal O}(\alpha)$ only, without exponentiation. Further progress has been made since and will continue steadily~\cite{Jadach:2018jjo}. For example, for the last three years in OPAL, the contribution of light fermion-pair production in the reaction ${\rm e^+e^- \to e^+e^-X}$ ${\rm (X=ee, \mu\mu, \dots)}$, giving rise to final states in configurations which would be accepted by the luminosity Bhabha selection, was evaluated according to Ref.~\cite{Montagna:1998vb,Montagna:1999eu} and corrected for in Ref.~\cite{Abbiendi:1999zx}. This correction reduced the OPAL luminosity uncertainty from 0.061\% to 0.054\%. A similar reduction can be contemplated for the other experiments. The vacuum polarization was recently
\begin{table}[htbp]
\small
\centering
\caption{Inspired from Refs.~\cite{Pietrzyk:1994rk, Jadach:1996gu, Jadach:2018jjo}: Summary of the theoretical uncertainties for a typical LEP luminosity detector covering the angular range from 58 to 110\,mrad (first generation) or from 30 to 50\,mrad (second generation). The total uncertainty is the quadratic sum of the individual components. 
\vspace{1mm}}
\label{tab:breakdown}
\begin{tabular}{|l|l|l|l|l|l|l|}
\hline
{\small LEP Publication in: } & \multicolumn{2}{|c|}{1994} & \multicolumn{2}{|c|}{2000} & \multicolumn{2}{|c|}{2019} \\ \hline
{\small LumiCal generation} & $1^{\rm st}$ & $2^{\rm nd}$ & $1^{\rm st}$ & $2^{\rm nd}$ & $1^{\rm st}$ & $2^{\rm nd}$ \\ \hline\hline
{\small Photonic ${\cal O}(\alpha^2 L_{\rm e})$} & 0.15\% &  0.15\% & 0.027\% & 0.027\% & 0.027\% & 0.027\% \\
{\small Photonic ${\cal O}(\alpha^3 L_{\rm e}^3)$} & 0.09\% & 0.09\% & 0.015\% & 0.015\% & 0.015\% & 0.015\% \\
{\small Z exchange} & 0.11\% & 0.03\% & 0.09\% & 0.015\% & 0.090\% & 0.015\% \\
{\small Vacuum polarization} & 0.10\% & 0.05\% & 0.08\% & 0.040\% & 0.015\% & 0.009\% \\
{\small Fermion pairs} & 0.05\% & 0.04\% & 0.05\% & 0.040\% & 0.010\% & 0.010\% \\ \hline
{\small Total} & 0.25\% & 0.16\% & 0.13\% & 0.061\% & 0.100\% & 0.037\% \\ \hline 
\end{tabular}
\end{table}
re-evaluated~\cite{Davier:2017zfy,Jegerlehner:2017zsb,Keshavarzi:2018mgv,Blondel:2019vdq,Davier:2019can}, which potentially further reduces the LEP luminosity uncertainty to 0.037\% in the second-generation LumiCals, and to 0.10\% in the first-generation LumiCals~\cite{Jadach:2018jjo}, as summarized in Table~\ref{tab:breakdown}.

The effects of these improvements on the integrated luminosities collected by the four LEP detectors are examined in turn in this letter. The Bhabha event selections are first recalled in Section~\ref{sec:selection}, and a brief account of their software emulation is given. The effect of the improved {\tt BHLUMI} prediction for the Z-exchange contribution is evaluated in Section~\ref{sec:Zexchange}, and that of the most recent estimate of the vacuum polarization in the $t$-channel photon propagator in Section~\ref{sec:VP}. Light fermion-pair production is studied in Section~\ref{sec:softpairs}. The combination of these corrections in a global fit of the four LEP experiments is described in Section~\ref{sec:combination}. A short summary is offered in Section~\ref{sec:conclusion}.

\section{Bhabha event selections}
\label{sec:selection}
As mentioned in Section~\ref{sec:introduction}, Bhabha events were selected at LEP by requiring the presence of two energy deposits (hereafter called ``clusters'') in the LumiCals. One cluster is required to be detected in the narrow polar-angle acceptance, as defined in Table~\ref{tab:AngularRanges}, while the other must be in the wide polar-angle acceptance on the other side of the interaction point. If more than two such clusters are found, only the most-energetic cluster is kept on each side. Let $E_{1,2}$, $\theta_{1,2}$, and  $\phi_{1,2}$ be the energies, polar angles, and azimuthal angles of these two clusters. The polar angle $\theta$ of the electron (positron) emerging from a Bhabha interaction is defined with respect to the direction of the $\rm e^-$ ($\rm e^+$) beam. The acollinearity and acoplanarity angle $\Delta\theta$ and $\Delta\phi$, defined by $\Delta\theta = \vert \theta_1-\theta_2 \vert$ and $\Delta\phi = \vert \vert \phi_1 - \phi_2 \vert - \pi \vert$, are expected to be vanishingly small for Bhabha events without non-soft initial state and final state radiation (ISR and FSR). 

The Bhabha selection criteria applied by the four LEP experiments to the clustered energies and to the acoplanarity/acollinearity angles are recalled in Table~\ref{tab:KinematicCuts}. To emulate these event selections in a quasi-realistic manner, imaginary detectors -- inspired from the study carried out in Ref.~\cite{Jadach:1996gu} -- are defined, consisting of a pair of cylindrical calorimeters, symmetrically located around the beam axis with respect to the interaction point, and covering the physical polar angle ranges of the actual LEP LumiCals. The beams are assumed to be point-like and to collide at the symmetry centre of the two calorimeters. These calorimeters are each divided into several azimuthal segments and radial pads, as in the actual detectors, such that the narrow and wide acceptance limits correspond to pad edges.\footnote{The ALEPH LCAL emulation is slightly different, as the pads are squares with vertical and horizontal edges. The wide and narrow acceptances follow these edges. The clustering process is also different, as it combines all energy deposits from each side into one cluster.} In a given event (Bhabha or otherwise), photons and electrons produced in the detector acceptance deposit their full energies in the pad they hit, while other particles (muons, neutrinos, pions, ...) are assumed to escape undetected. No shower simulation is attempted, and no energy smearing is applied.
\begin{table}[!htbp]
\small
\centering
\caption{Kinematic selection criteria applied to the clustered energies $E_{1,2}$ deposited in the two LumiCals, and on the acoplanarity and acollinearity angles between the two clusters, $\Delta\phi$ and $\Delta\theta$. The beam energy is denoted $E_{\rm beam}$. Some of the selection criteria changed during the LEP\,1 era. The periods of validity are also indicated. From 1993 onward in L3, the smaller cluster energy was allowed to be as small as 20\% of the beam energy if the larger one energy exceeded 95\% of the beam energy, in order to recover events with energy lost in the gaps between crystals. Also, in the first generation LumiCals (and in L3 over the whole LEP1 period), the clusters were required to be away from the vertical separation between the two halves of the calorimeters. Because the imaginary detectors considered here have no gaps and cracks, these last cuts are not emulated. This choice does not affect the relative cross-section changes studied in this letter. %
\vspace{2mm} }
\label{tab:KinematicCuts}
\begin{tabular}{|l|c|c|c|c|}
    \hline
    \multirow{2}{*}{Experiment}     &   
    ALEPH & DELPHI & L3 &  OPAL  \\
    & \cite{Barate:1999ce} &
    \cite{Abreu:258291,Abreu:259634,Abreu:2000mh} &
    \cite{Acciarri:261072,Acciarri:2000ai} &
    \cite{Acton:245031,Akers:252425,Abbiendi:2000hu} \\
    \hline\hline
    $E_{1,2}^{\rm min}/E_{\rm beam}$ &  \multirow{ 2}{*}{$> 0.44$} & \multirow{ 2}{*}{$> 0.65$} & $> 0.40$ &  $> 0.45$ ($\to$ 92) \\ 
    $E_{1,2}^{\rm max}/E_{\rm beam}$ &  &  & $> 0.80$ &  $> 0.50$ (93 $\to$) \\ \hline
                              &  $> 0.60$ ($\to$ 93) & \multirow{3}{*}{--} & \multirow{3}{*}{--} &  \multirow{2}{*}{$>0.67$ ($\to$ 92)} \\
    $\displaystyle \frac{(E_1+E_2)}{2E_{\rm beam}}$ &  $> 0.78$ (in 94) & & &  \multirow{2}{*}{$> 0.75$ (93 $\to$)} \\
                              &  $> 0.84$ (in 95) & & &  \\ \hline
    \multirow{2}{*}{$\Delta \phi$ (mrad)} &  $< 175$ ($\to$ 8/92) & \multirow{2}{*}{$< 350$} & \multirow{2}{*}{$< 175$} & $<350$ ($\to$ 92) \\ 
                         &  $< 525$ (9/92 $\to$) &  & &  $< 200$ (93 $\to) $       \\ \hline
    \multirow{2}{*}{$\Delta \theta$ (mrad)} & \multirow{2}{*}{--} & \multirow{2}{*}{--} & \multirow{2}{*}{--} &  \ \ \ -- \ \  ($\to$ 92)\\ 
                           & &  &  &  $< 10$ (93 $\to$) \\ \hline
\end{tabular}
\end{table}

The pad with the most energetic deposit defines a seed, and the energies detected in pads sufficiently close to the seed (where the ``sufficiently close'' criterion depends on the LumiCal at stake) are combined into a cluster. The cluster energy is the sum of all pad energies, and the coordinates of the cluster are the energy-weighted averaged ($\theta, \phi$) coordinates, summing over all pads in the cluster. Events are selected as displayed in Table~\ref{tab:KinematicCuts} using the cluster energies and angles on both sides. The emulation reproduces with a reasonable accuracy the reference cross sections at the Z peak, as published by the experiments. For example, it predicts cross sections of 84.48\,nb and 78.74\,nb for ALEPH and OPAL in their 1994 configurations, to be compared with the published references of $\sim 84$\,nb~\cite{Barate:1999ce} and $78.71$\,nb~\cite{Abbiendi:2000hu}, respectively. The absolute cross-section value is, however, of moderate importance for the present study, as only relative changes are evaluated hereafter.


\section{The Z-exchange contribution}
\label{sec:Zexchange}

\begin{figure}[htbp]
\centering
\includegraphics[width=0.42\columnwidth]{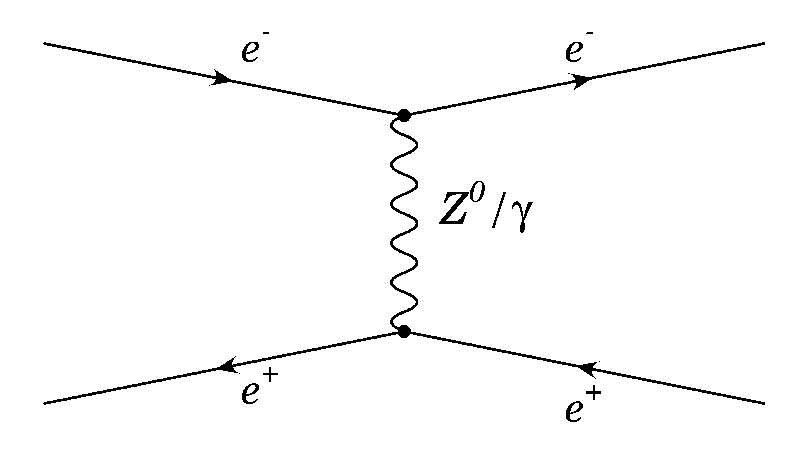} 
\includegraphics[width=0.42\columnwidth]{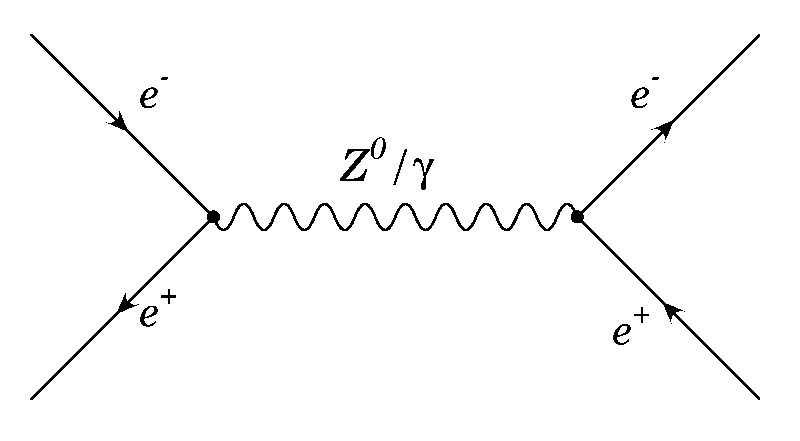}
\caption{\small Born approximation of the low-angle Bhabha scattering, with photon and Z exchange in the $t$ (left) and $s$ (right) channels, 
as included in {\tt BHLUMI 2.01}
}
\label{fig:Zexchange}
\end{figure}

In {\tt BHLUMI 2.01}, used by ALEPH, DELPHI, and L3 until the end of 1992, the Z-exchange contribution to low-angle Bhabha scattering was included in the Born approximation only. The corresponding Feynman diagrams are displayed in Fig.~\ref{fig:Zexchange}. The ${\cal O}(\alpha)$ QED corrections to this contribution, included in {\tt BABAMC} are sizeable (up to $\sim 50\%$) because of its strong energy dependence around the Z pole. 
The {\tt BHLUMI} events simulated at the time in ALEPH LCAL, DELPHI SAT, L3 BGO, and ALEPH SiCAL'92 were therefore reweighted by the {\tt BABAMC} estimate of this correction. 

In {\tt BHLUMI 4.0x}, the QED corrections to the Z contribution have been implemented in several manners in the form of auxiliary weights, of which {\it (i)} an emulation of the {\tt BABAMC} ${\cal O}(\alpha)$ version, checked to reproduce well the {\tt BABAMC} prediction; and {\it(ii)} a new version with the ${\cal O}(\alpha)$ YFS exponentiation of Ref.~\cite{Jadach:1995hv} that includes higher-order effects. (An auxiliary weight for the Z contribution as implemented in {\tt BHLUMI 2.01} is also available.) The latter, more accurate, implementation was used by all experiments from 1993 onward, and used by OPAL to rescale their FD luminosity measurements until 1992 in their last publication~\cite{Abbiendi:2000hu}. 

\begin{figure}[htbp]
\centering
\begin{tabular}{cc}
\includegraphics[width=0.70\columnwidth]{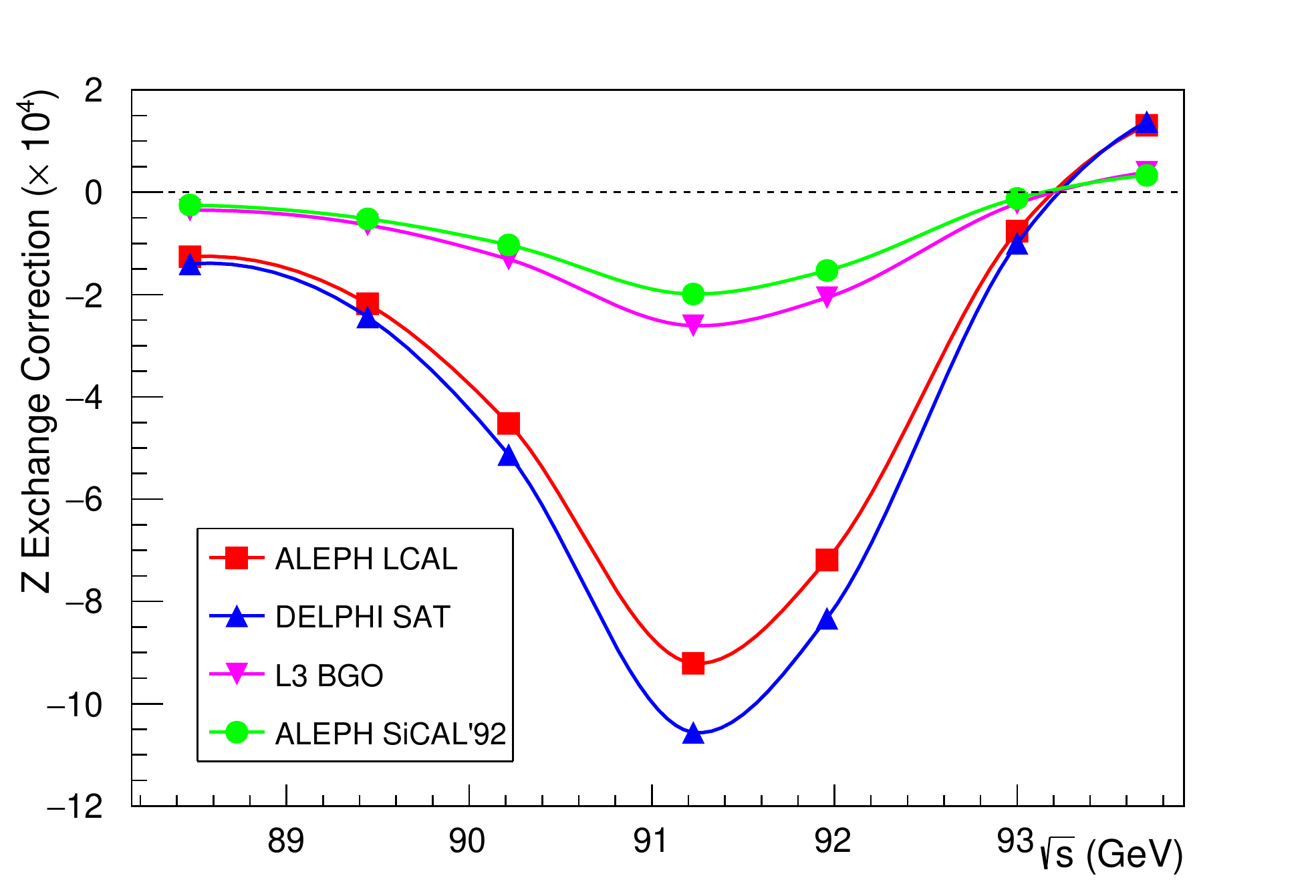} 
\end{tabular}
5\caption{\small Relative Bhabha cross-section change in ALEPH LCAL, DELPHI SAT, L3 BGO and ALEPH SiCAL'92, in units of $10^{-4}$, when using the most accurate estimate of the QED corrections to the Z-exchange contribution, as implemented in {\tt BHLUMI 4.04}, instead of the estimate that of {\tt BABAMC} used until 1992.}
\label{fig:zrescale}
\end{figure}

These two implementations are still accessible in the current version of {\tt BHLUMI}. Over one billion Bhabha events were therefore generated with {\tt BHLUMI 4.04} with weights corresponding to each of these two Z-exchange estimates, and simulated in ALEPH LCAL, DELPHI SAT, L3 BGO, and ALEPH SiCAL, as explained in Section~\ref{sec:selection}. The relative difference between the two estimates is displayed in Fig.~\ref{fig:zrescale} and in Table~\ref{tab:zrescale}, as a function of centre-of-mass energy at and around the Z peak. (The changes in OPAL FD are not included for the reasons explained above.)

\begin{table}[htbp]
\small
\centering
\caption{Relative Bhabha cross-section change in ALEPH LCAL, DELPHI SAT, L3 BGO and ALEPH SiCAL'92, in units of $10^{-4}$, when the {\tt BHLUMI 4.04} YFS-exponentiated QED-corrections to the Z-exchange contribution are used instead of the {\tt BABAMC} ${\cal O}(\alpha)$ estimate available until 1992, as a function of centre-of-mass energy $\sqrt{s}$. The $\sqrt{s}$ values correspond to the LEP luminosity-weighted averages. The uncertainties are statistical. %
\vspace{2mm} }
\label{tab:zrescale}
\begin{tabular}{|c|r|r|r|r|}
    \hline
    $\sqrt{s}$ (GeV) &   
    {\small ALEPH LCAL} & {\small DELPHI SAT} & {\small L3 BGO} &  {\small ALEPH SiCAL}  \\ \hline\hline
    88.471 & $-1.26 \pm 0.03$ & $-1.40 \pm 0.03$ & $-0.35 \pm 0.03$ & $ -0.25 \pm 0.05$ \\
    89.444 & $-2.18 \pm 0.04$ & $-2.44 \pm 0.04$ & $-0.64 \pm 0.01$ & $ -0.52 \pm 0.01$ \\
    90.216 & $-4.52 \pm 0.05$ & $-5.13 \pm 0.05$ & $-1.31 \pm 0.01$ & $ -1.03 \pm 0.01$ \\
    91.227 & $-9.21 \pm 0.03$ & $-10.56 \pm 0.04$ & $-2.61 \pm 0.01$ & $-1.99 \pm 0.01$ \\
    91.959 & $-7.19 \pm 0.05$ & $-8.33 \pm 0.05$ & $-2.06 \pm 0.01$ & $ -1.53 \pm 0.01$ \\
    93.000 & $-0.76 \pm 0.05$ & $-1.00 \pm 0.05$ & $-0.22 \pm 0.01$ & $ -0.12 \pm 0.01$ \\
    93.710 & $+1.31 \pm 0.05$ & $+1.38 \pm 0.05$ & $+0.39 \pm 0.01$ & $ +0.33 \pm 0.01$ \\
    \hline
\end{tabular}
\end{table}

At the Z peak, the updated Bhabha cross section is up to $0.1\%$ smaller than that used for the published results, for ALEPH LCAL and DELPHI SAT, both with an inner polar angle of $\sim 56$\,mrad. The difference is five times less in L3 BGO and ALEPH SiCAL'92, because of the smaller inner polar angle of $\sim 31$\,mrad, therefore yielding a much reduced Z-exchange absolute contribution. These changes are well within the 1994 theoretical uncertainties of $\pm 0.11\%$ (large angle) and $\pm 0.03\%$ (small angle) quoted in Table~\ref{tab:breakdown}. The new theoretical uncertainties are evaluated to be $\pm 0.09\%$ and $\pm 0.015\%$, respectively.

\section{Vacuum Polarization}
\label{sec:VP}

\begin{figure}[htbp]
\centering
\begin{tabular}{cc}
\includegraphics[width=0.20\columnwidth]{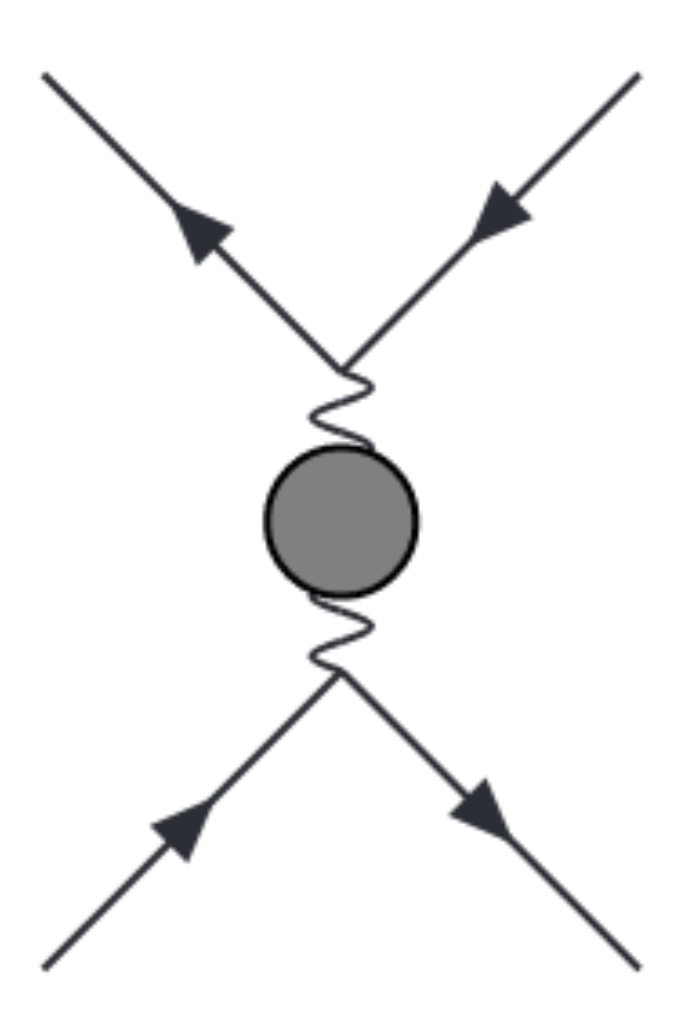} 
\end{tabular}
\caption{\small Vacuum polarization in the $t$-channel photon propagator of the small-angle Bhabha scattering.}
\label{fig:vacpol}
\end{figure}

In small-angle Bhabha scattering, vacuum polarization appears in the $t$-channel photon propagator dressed with a loop of leptons or quarks, as illustrated in Fig.~\ref{fig:vacpol}. The leptonic contribution is now known to up to four loops~\cite{Steinhauser:1998rq,Sturm:2013uka}, i.e. with virtually infinite precision. The hadronic part is obtained with a combination of experimental cross-section data involving ${\rm e^+ e^-}$ annihilation to hadrons (called $R$ ratio), and kernels that can be calculated with perturbative QCD. 

Significant progress was achieved in the past two decades on the experimental determination of the $R$ ratio as a function of the centre-of-mass energy, in particular with recent analyses of the huge samples of hadronic events with ISR collected at $\Phi$ and B factories. A summary of the present situation can be found in Refs.~\cite{Davier:2017zfy,Jegerlehner:2017zsb,Keshavarzi:2018mgv,Blondel:2019vdq,Davier:2019can}. The December 2019 version of the momentum-transfer-dependent code of Jegerlehner~\cite{Jegerlehner:hadr5x} is used in this letter to evaluate the corresponding changes in the low-angle Bhabha scattering cross section. The results were cross-checked and confirmed with independent codes, developed by the KNT~\cite{Keshavarzi:2018mgv} and DHMZ~\cite{Davier:2019can} teams, and updated in February 2020.

The same one billion Bhabha events as in Section~\ref{sec:Zexchange} were used for this purpose, generated with weights corresponding to the vacuum polarization evaluated in 1988~\cite{Burkhardt:201068} (implemented in {\tt BHLUMI 2.0x}), in 1995~\cite{Burkhardt:1995tt, Eidelman:1995ny} (implemented in {\tt BHLUMI 4.0x}), and in 2019~\cite{Jegerlehner:hadr5x}. The relative cross-section modifications (which also include -- when relevant -- non-soft ${\cal O}(\alpha^2 L_{\rm e}^2)$ QED corrections, implemented in {\tt BHLUMI 4.04}) are displayed in Table~\ref{tab:VacuumPolarization}. The updated cross section appear to be smaller than that used for the LEP experiment publications, by about $-0.022\%$ for the second-generation LumiCals, and in the $(-0.015 \pm 0.030)\%$ range for the first generation LumiCals, well within the systematic uncertainties evaluated at the time, of $\pm 0.04$-$0.05$\% and $\pm 0.08$-$0.10$\%, respectively. After correction, the new theoretical uncertainties are estimated to be of the order of $\pm 0.009\%$ and $\pm 0.015\%$.

{\renewcommand{\arraystretch}{1.3}
\begin{table}[htbp]
\small
\centering
\caption{Vacuum polarization correction relative to the Bhabha cross section at the Z peak (switching off the Z boson contribution) in the LumiCals of the four LEP experiments, as a function of time. Each entry value, expressed in units of $10^{-4}$, corresponds to the change between the 2019 evaluation of the vacuum polarization in the $t$-channel photon propagator, and the default function proposed in the {\tt BHLUMI} versions used in the LEP experiments' publications at the time. When relevant, the entries also include non-soft ${\cal O}(\alpha^2 L_{\rm e}^2)$ corrections implemented only in {\tt BHLUMI 4.0}. Statistical uncertainties are negligible and are not quoted in the table. The numbers display the quasi-linear excursion of the correction when varying $\sqrt{s}$ from $88.471$\,GeV (subscript) to $93.710$\,GeV (superscript).  \vspace{0mm}}
\label{tab:VacuumPolarization}
\begin{tabular}{|r|>{\hspace{2mm}}c<{\hspace{2mm}}|>{\hspace{2mm}}c<{\hspace{2mm}}|>{\hspace{2mm}}c<{\hspace{2mm}}|>{\hspace{2mm}}c<{\hspace{2mm}}|}
\hline 
Experiment  &  ALEPH & DELPHI & L3 & OPAL \\ \hline\hline
%
{\small 01/90 $\to$ 08/92} & $-2.00^{-0.18}_{+0.21}$ & \multirow{2}{*}{$-1.02^{-0.18}_{+0.18}$} & \multirow{2}{*}{$+1.57^{-0.09}_{+0.10}$} & \multirow{2}{*}{$-4.60^{-0.03}_{+0.04}$} \\ \cline{1-2}
{\small 09/92 $\to$ 12/92} & $+1.22^{-0.12}_{+0.12}$ & & & \\ \cline{1-5} 
{\small 01/93 $\to$ 12/93} & \multirow{3}{*}{$-2.12^{-0.08}_{+0.09}$} & $-4.62^{-0.06}_{+0.07}$ & \multirow{3}{*}{$-2.36^{-0.09}_{+0.11}$} & \multirow{3}{*}{$-2.24^{-0.09}_{+0.10}$} \\ \cline{1-1} \cline{3-3}
{\small 01/94 $\to$ 12/94} & & \multirow{2}{*}{$-3.86^{-0.11}_{+0.12}$} & & \\ \cline{1-1}
{\small 01/95 $\to$ 12/95} & & & & \\ \hline\hline
\end{tabular}
\end{table}
}

\section{Light fermion-pair production}
\label{sec:softpairs}

\begin{figure}[htbp]
\centering
\begin{tabular}{cc}
\begin{minipage}[b]{0.60\textwidth}
\centering
\hspace{-0.75cm}
\includegraphics[width=\textwidth]{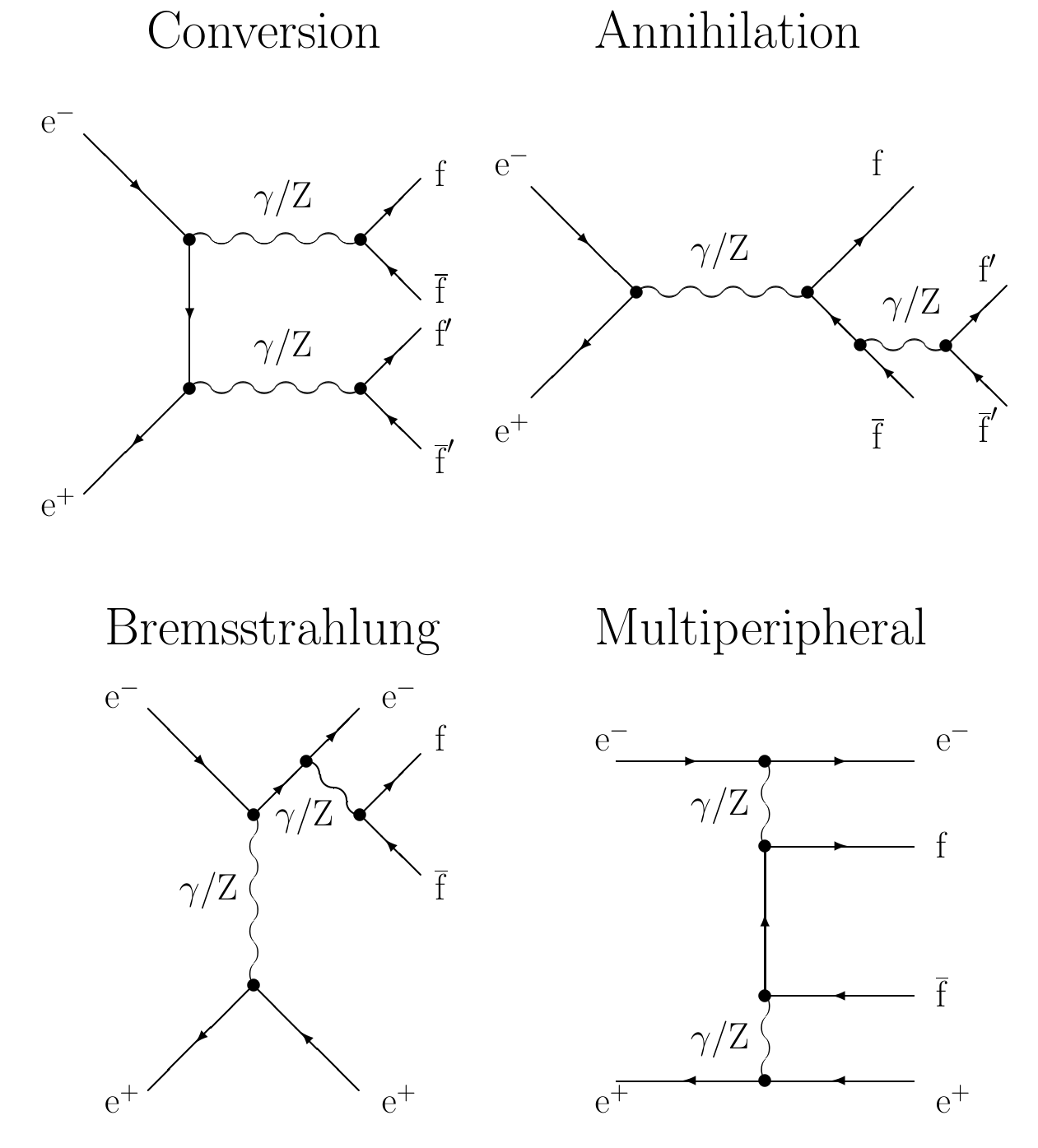}
\end{minipage}
\begin{minipage}[b]{0.28\textwidth}
\centering
{Virtual} 

\vspace{0.4cm} 
\includegraphics[width=\textwidth]{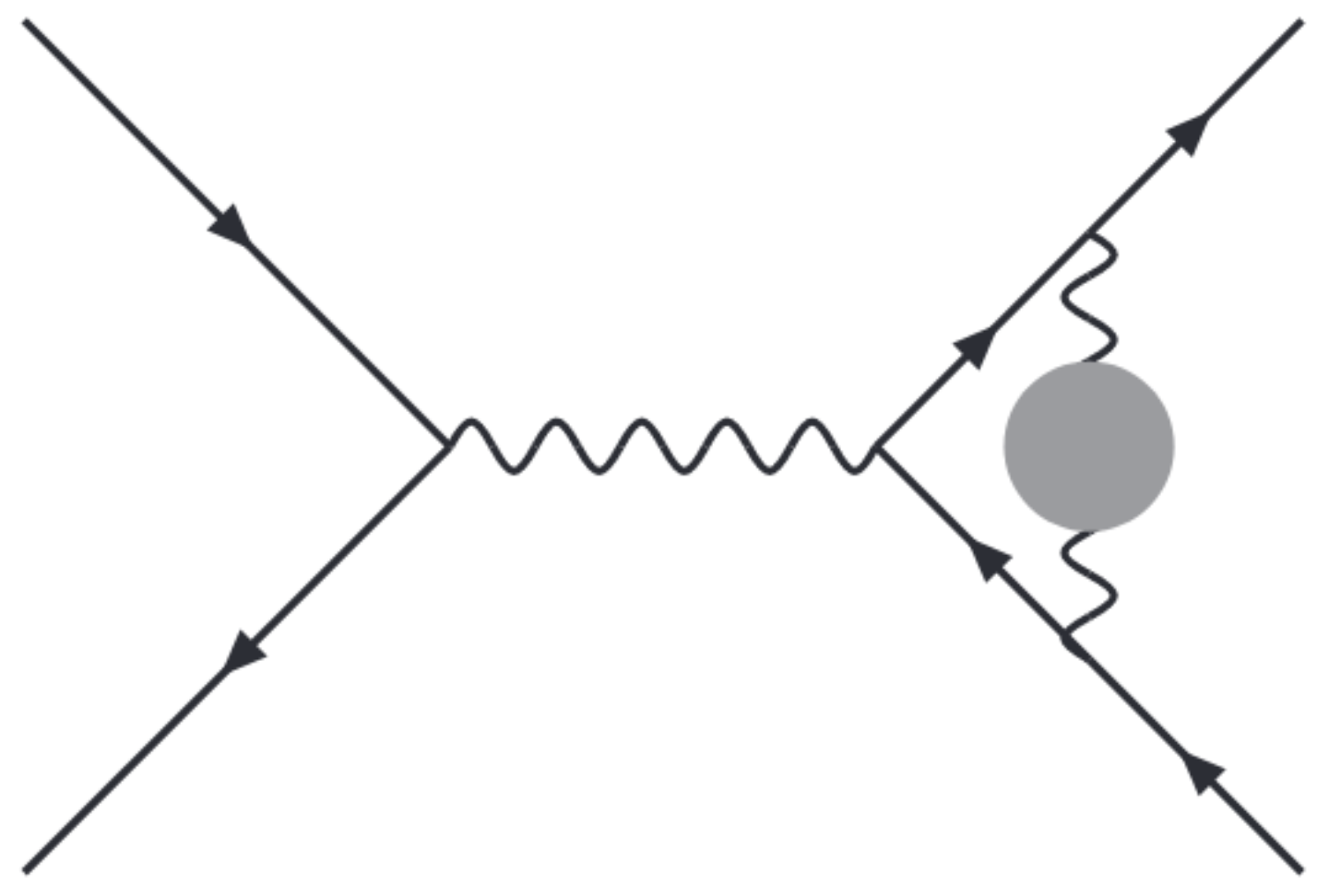}

\vspace{0.8cm}
\includegraphics[width=0.7\textwidth]{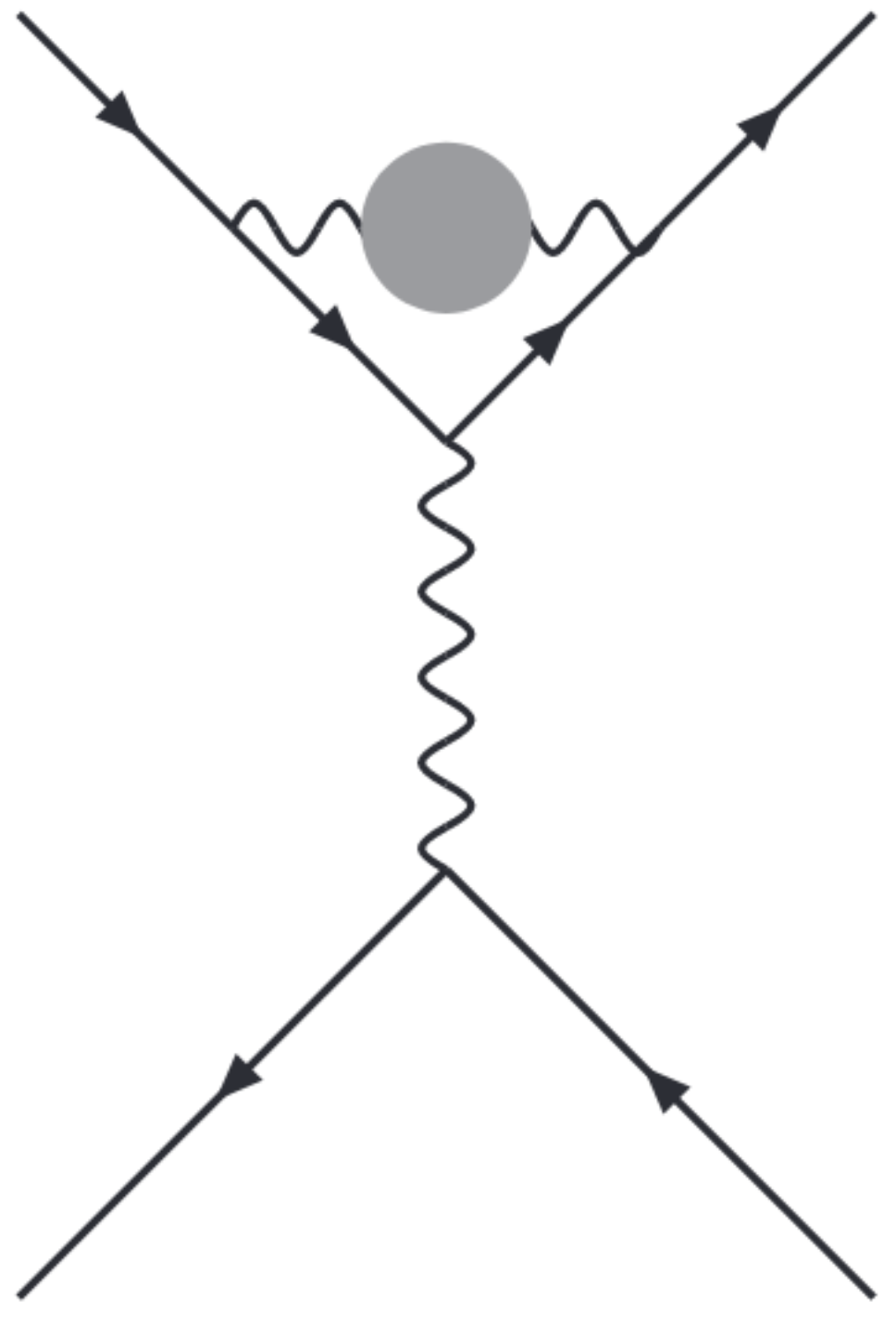} 

\vspace{0.4cm}
\end{minipage}
\end{tabular}
\caption{\small (Left) Selection of lowest-order diagrams for the ${\rm e^+e^- \to e^+e^- f \bar{f}}$ four-fermion process; (Right) Virtual vertex correction to the Bhabha scattering process with the corresponding one-loop fermion-pair insertion (grey circle). 
}
\label{fig:SoftPair}
\end{figure}

The next outstanding theoretical contributions to the uncertainty on the Bhabha cross section used by the LEP experiments is the effect of additional fermion-pair production. 
Additional fermion-pair production arises from the ${\rm e^+e^- \to e^+e^- f \bar{f}}$ four-fermion process, where ${\rm f}$ is either a charged lepton, or a quark, or a neutrino, and which gives rise to final states in configurations that can be accepted by the luminosity Bhabha selection. To estimate this effect with precision, both the real production (with a real ${\rm f \bar{f}}$ pair in the final state) and the virtual correction to the ${\rm e^+e^- \to e^+e^-}$ process (with a virtual ${\rm f \bar{f}}$ loop off the external ${\rm e^\pm}$ lines, as displayed in Fig.~\ref{fig:SoftPair}) have to be included. 

Such estimates exist in the literature, either in the ALEPH LCAL acceptance~\cite{Jadach:320449} or in the OPAL SiW acceptance~\cite{Montagna:1998vb, Montagna:1999eu}, with some approximations. The former considers just the leading correction, with only ${\rm f = e}$ in the $t$-channel approximation, and omits the $175$\,mrad acoplanarity cut of Table~\ref{tab:KinematicCuts}. The latter includes all charged leptons (but not the quarks), all Feynman diagrams of Fig.~\ref{fig:SoftPair}, and all OPAL selection criteria. This estimate was used by the OPAL Collaboration as a correction to the reference Bhabha cross section from {\tt BHLUMI 4.04} to determine their 1993-1995 integrated luminosity in their final publication on the matter~\cite{Abbiendi:1999zx}. These two pioneering analyses are generalised in this letter to the eight angular ranges of Table~\ref{tab:AngularRanges}, the eight sets of selection criteria of Table~\ref{tab:KinematicCuts} (four for ALEPH, two for OPAL, and one for DELPHI and L3), eight fermion flavours for the virtual and real contributions (top-quark and neutrino contributions are negligible), and all lowest-order four-fermion diagrams of Fig.~\ref{fig:SoftPair} for the real contribution. 

The real contribution is evaluated  for each fermion 
${\rm f}$ with the {\tt FERMISV} Monte Carlo program~\cite{Hilgart:1992xu}, which includes all neutral-current four-fermion diagrams (Fig.~\ref{fig:SoftPair}), ISR to first order, and FSR from all charged fermions. The contribution of final states with quarks is corrected as explained in Ref.~\cite{Buskulic:271816} by the value of the ratio of the experimental ${\rm e^+e^- \to}$ hadrons cross section to the prediction of the quark-parton model, trivially related to the $R$ ratio. In the generation of all ${\rm e^+e^- f\bar{f}}$ final states, the more energetic electron (positron) must be in the LumiCal 
acceptance, on the side of the interaction point defined by the electron (positron) beam direction. For the ${\rm e^+e^-e^+e^-}$ final state, the other two ${\rm e}^\pm$ are required to be at least $0.1^\circ$ away from the beam axis, to avoid a pole of the cross section that is not efficiently regularized in {\tt FERMISV}. This pole is adequately regularized all the way to $0^\circ$ in the {\tt KORALW} Monte Carlo program~\cite{Jadach:1998gi}, which is therefore used to correct the {\tt FERMISV} ${\rm e^+e^-e^+e^-}$ cross section for the missing phase space.\footnote{On the other hand, the {\tt KORALW} version used for this work develops large weight fluctuations when the first ${\rm e^+e^-}$ pair is at low angle, and therefore could not be used to determine the absolute cross section value with an acceptable statistical accuracy in a reasonable time. In a phase space region with the first ${\rm e^+e^-}$ pair above $3^\circ$ and the second pair above $0.1^\circ$, the {\tt FERMISV} and {\tt KORALW} predicted cross sections agree within a couple~\%.} This correction is at the 5-to-10\% level as per the LumiCal acceptance and the selection criteria.

The virtual contribution is determined at the one-loop level by a simple Monte Carlo simulation code with, as suggested in Ref.~\cite{Actis:2008br}: 
\begin{equation}
\begin{aligned}
    \frac{{\rm d}\sigma(s,t)}{{\rm d}\Omega} = \frac{\alpha^2}{2s}  & \left\{ T_s  \left( 1 + 4\frac{\alpha^2}{\pi^2}\Re V_2(s) \right) +
    T_t \left( 1 + 4\frac{\alpha^2}{\pi^2} V_2(t) \right) \right. \\ 
    & \left. + T_{st}  \left( 1 + 2\frac{\alpha^2}{\pi^2} \left[ \Re V_2(s)+ V_2(t) \right] \right)\right\}, 
\end{aligned}
\end{equation}
where $T_s$, $T_t$, and $T_{st}$ are the $s$-channel, $t$-channel, and interference terms of the lowest-order differential Bhabha cross section at a given ${\rm e^+e^-}$ centre-of-mass energy $\sqrt{s}$ and for a given momentum transfer squared $t$ (with both Z and $\gamma$ exchanges); and the $V_2(x)$ terms ($x = s, t$), which include the QED corrections from virtual fermion loops, are taken from Ref.~\cite{Actis:2008br}. Here again, the hadronic part is corrected for the value of the $R$ ratio. Initial-state radiation is dealt with by an upgraded version of the {\tt REMT} package~\cite{Berends:1984dw} allowing for up to two ISR photons.

Four-fermion and Bhabha events are then processed and selected as described in Section~\ref{sec:selection}. The virtual contribution comes out as a negative correction to the lowest-order ISR-corrected Bhabha cross section, providing a delicate cancellation with the real contribution when taken inclusively. When the kinematic selection criteria of  Table~\ref{tab:KinematicCuts} are applied, however, the sum of the real and virtual contributions results in a small, negative, correction to the Bhabha cross section: the tighter the selection, the larger the correction in absolute value. These corrections are summarized in Table~\ref{tab:SoftPairCorrection} for each LEP experiment as a function of time. They vary from $-0.03\%$ to $-0.05\%$.

\begin{table}[htbp]
\small
\centering
\caption{Light fermion-pair correction relative to the Bhabha cross section selected by the LumiCals of the four LEP experiments, as a function of time. Entry values sum up real and virtual corrections, and are expressed in units of $10^{-4}$. The corrections are found to be independent of the centre-of-mass energy (within statistical uncertainties). The indicated uncertainties combine statistical uncertainties and an estimate of the systematic effect of the detector granularity and the clustering procedure. The latter is taken to be equal to the half the difference between the correction obtained with the detector/clustering emulation and that obtained with the electron and positron exact energies and directions.
\vspace{0mm}}
\label{tab:SoftPairCorrection}
\begin{tabular}{|r|c|c|c|c|}
\hline 
Experiment  &  ALEPH & DELPHI & L3 & OPAL \\ \hline\hline
{\small 01/90 $\to$ 08/92} & $-3.58 \pm 0.06$ & \multirow{3}{*}{$-4.99 \pm 0.06$} & \multirow{2}{*}{$-3.43 \pm 0.04$} & \multirow{2}{*}{$-4.51 \pm 0.09$} \\ \cline{1-2}
{\small 09/92 $\to$ 12/92} & \multirow{2}{*}{$-3.00 \pm 0.06$} & & & \\ \cline{1-1} \cline{4-5}
{\small 01/93 $\to$ 12/93} & & & \multirow{3}{*}{$-3.77 \pm 0.07$} & $-4.72 \pm 0.17$ \\ \cline{1-3}
{\small 01/94 $\to$ 12/94} & $-3.52 \pm 0.08$ & \multirow{2}{*}{$-3.91 \pm 0.05$} &  & {\footnotesize ($-4.40$ already}\\ \cline{1-2}
{\small 01/95 $\to$ 12/95} & $-4.38 \pm 0.08$ & & & {\footnotesize applied in~\cite{Abbiendi:1999zx})} \\ \hline\hline
\end{tabular}
\end{table}

To test the technical validity of the procedure, the results presented in Table~\ref{tab:SoftPairCorrection} were compared to previous estimates~\cite{Jadach:320449,Montagna:1999eu}. When only the ${\rm f = e}$ contribution is included in the Bremsstrahlung four-fermion diagrams and in the Bhabha $t$-channel photon-exchange term, and only the energy cuts are applied, the real and virtual corrections cancel almost exactly: their sum amounts to $(-0.47 \pm 0.12) \times 10^{-4}$ in the ALEPH LCAL acceptance. With a similar setup, a correction of $(-1.0 \pm 0.5) \times 10^{-4}$ had been obtained -- both analytically and with Monte Carlo simulation -- in Ref.~\cite{Jadach:320449}. Similarly, when all charged lepton (${\rm e}, \mu, \tau$) contributions are included (but no quark contribution), the sum of the real and virtual correction amounts to $(-4.18 \pm 0.17) \times 10^{-4}$ in the OPAL SiW acceptance, with all selection criteria applied. This value is well compatible with the $(-4.4 \pm 1.4) \times 10^{-4}$ obtained in Ref.~\cite{Montagna:1999eu} with the same experimental setup, but an entirely different Monte Carlo suite.

These two successful tests allow the technical accuracy of the procedure to be evaluated from the difference between the various estimates, at the level of $\pm 0.2 \times 10^{-4}$. This technical uncertainty is inflated to $\pm 0.6 \times 10^{-4}$ to account for the limitations of {\tt FERMISV} and {\tt KORALW} in the vicinity of the beam axis for the ${\rm e^+e^-e^+e^-}$ final state. An uncertainty of 20\%, i.e., $\pm 0.8 \times 10^{-4}$, is assigned to the contribution of higher-order corrections, leading to a total theoretical uncertainty of about $\pm 0.01\%$.

\section{Combined fit} 
\label{sec:combination}

Remarkably, each and every correction presented in the previous sections tends to decrease the Bhabha cross section with respect to that used by the LEP experiments in their publications. This smaller cross section increases the integrated luminosity, thus decreases the hadronic cross section at all centre-of-mass energies, and leads in turn to a larger number of light neutrino species (Eq.~\ref{eq:NnuForLep}). The exact shift is evaluated as explained in this section. 

At LEP, data were produced at centre-of-mass energies values from 88 to 94\,GeV, clustered around seven values called ``Peak$-$3'', ``Peak$-$2'', ..., Peak, ..., `Peak$+$2'', and `Peak$+$3''. The luminosity-weighted averages of these centre-of-mass energies are indicated in Table~\ref{tab:zrescale}. In a first step, the luminosity increase due to the three modified contributions to the Bhabha cross section (Z exchange, vacuum polarization, and light fermion-pairs) is therefore calculated, at each of these seven centre-of-mass energies and for each experiment, as the error-weighted average  of the biases displayed in Tables~\ref{tab:zrescale} to~\ref{tab:SoftPairCorrection}.  This average is taken over the 1990 to 1995 period, accounting for statistical uncertainties and year-to-year uncorrelated systematic uncertainties on the recorded luminosities. In this average, the 1990 uncertainties are artificially inflated by a large factor because {\it (i)} OPAL increased their 1990 luminosity errors too~\cite{Abbiendi:2000hu}; {\it (ii)} ALEPH estimated their 1990 luminosities with {\tt BABAMC} instead of {\tt BHLUMI}; and {\it (iii)} the resonant depolarization method necessary to precisely determine the beam energy was not yet available, making the 1990 data not fully appropriate for an accurate lineshape~fit. 

Between 1991 and 1995, 98.5\% of the LEP data were produced at the Peak$-$2, Peak, and Peak$+$2 centre-of-mass energies. The small amount of  data recorded in 1991 at Peak$-$3 and Peak$-$1 on one hand, and at Peak$+$1 and Peak$+$3 on the other, are thus combined and merged with the Peak$-$2 and Peak$+$2 data, respectively, still with an error-weighted average. This simplification does not change the final result in any relevant manner. The outcome of these first two steps is displayed in Table~\ref{tab:LEPLuminosityBiases}. For completeness, the beam-induced effects of Ref.~\cite{Voutsinas:2019hwu}, and the total integrated-luminosity increase with respect to that of the last LEP combination~\cite{ALEPH:2005ab}, 
\begin{table}[htbp]
\small
\centering
\caption{Integrated-luminosity relative increase with respect to Ref.~\cite{ALEPH:2005ab}, determined for each of the four LEP experiments at the Peak$-$2, Peak, and Peak$+$2 centre-of-mass energies, due to the updated evaluations of the Z-exchange, the vacuum polarization, and the fermion-pair production contributions. The beam-induced luminosity increase~\cite{Voutsinas:2019hwu}, as well as the sum of all effects, are also indicated. All entries are in units of $10^{-4}.$\vspace{2mm}}
\label{tab:LEPLuminosityBiases}
{\bf Peak$-$2:}
\begin{tabular}{|l|r|r|r|r|}
\hline 
Source / Experiment          &  \phantom{D}ALEPH   & DELPHI & \phantom{DELP}L3 & \phantom{DE}OPAL \\\hline\hline
Z exchange                            & $ 0.10$ & $ 0.12$ & $ 0.03$ & $ 0.00$ \\ \hline
Light fermion-pairs                   & $ 3.39$ & $ 4.33$ & $ 3.76$ & $ 0.36$ \\ \hline
Vacuum polarization                   & $2.05$ & $3.90$ & $2.13$ & $2.20$  \\ \hline
Beam-induced~\cite{Voutsinas:2019hwu} & $ 8.77$ & $ 4.67$ & $ 8.36$ & $ 8.74$ \\\hline\hline
Total                                 & $14.30$ & $13.03$ & $14.27$ & $11.30$ \\ \hline
\end{tabular}
\\ \vspace{2mm} 
{\bf Peak\phantom{$-$2}:}
\begin{tabular}{|l|r|r|r|r|}
\hline 
Source / Experiment          &  \phantom{D}ALEPH   & DELPHI & \phantom{DELP}L3 & \phantom{DE}OPAL \\\hline\hline
Z exchange                            & $ 0.52$ & $ 0.35$ & $ 0.06$ & $ 0.00$ \\ \hline
Light fermion-pairs                   & $ 3.35$ & $ 4.07$ & $ 3.76$ & $ 0.40$ \\ \hline
Vacuum polarization                   & $1.82$ & $3.85$ & $2.28$ & $2.28$ \\ \hline
Beam-induced~\cite{Voutsinas:2019hwu} & $10.29$ & $ 5.67$ & $ 9.60$ & $10.55$ \\\hline\hline
Total                                 & $15.98$ & $13.94$ & $15.69$ & $13.24$ \\ \hline
\end{tabular}
\\ \vspace{2mm}
{\bf Peak$+$2:}
\begin{tabular}{|l|r|r|r|r|}
\hline 
Source / Experiment          &  \phantom{D}ALEPH   & DELPHI & \phantom{DELP}L3 & \phantom{DE}OPAL \\\hline\hline
Z exchange                            & $ 0.03$ & $ 0.03$ & $ 0.01$ & $ 0.00$ \\ \hline
Light fermion-pairs                   & $ 3.39$ & $ 4.19$ & $ 3.76$ & $ 0.36$ \\ \hline
Vacuum polarization                   & $2.18$ & $4.01$ & $2.27$ & $2.33$ \\ \hline
Beam-induced~\cite{Voutsinas:2019hwu} & $ 8.44$ & $ 4.58$ & $ 8.04$ & $ 8.40$ \\\hline\hline
Total                                 & $14.04$ & $12.81$ & $14.07$ & $11.10$ \\ \hline
\end{tabular}
\end{table}
are also indicated.\footnote{The luminosity biases due to beam-induced effects indicated here are slightly different from those of Table~4 in Ref.~\cite{Voutsinas:2019hwu} because the 1990-1992 data were not included therein.} 

Each contribution to the luminosity increase is accompanied by a reduction of the theoretical uncertainty as indicated in Table~\ref{tab:breakdown}. At each step, the nine Z lineshape parameters -- $m_{\rm Z}$, $\Gamma_{\rm Z}$, $\sigma^0_{\rm had}$, $R^0_{{\rm e},\mu,\tau}$ and $A_{\rm FB}^{0,({\rm e},\mu,\tau)}$ -- and the $9\times 9$ covariance matrix, are re-evaluated accordingly. Starting from the parameters and covariance matrices given in Table~2.4 and Table~2.9 of Ref.~\cite{ALEPH:2005ab} for each experiment, the errors on $m_{\rm Z}$, $\Gamma_{\rm Z}$, and $\sigma^0_{\rm had}$ are propagated back to the hadronic cross sections at the Peak$-$2, Peak, and Peak$+$2 centre-of-mass energies assuming a Breit-Wigner resonance; these hadronic cross sections and their uncertainty are reduced according to the corrected integrated luminosity; the new Breit-Wigner parameters are fit, and updated covariance matrices are obtained, both with and without assuming lepton universality; and the operation is repeated for each contribution to the luminosity increase.     

The updated Z parameters and covariance matrices, obtained after correcting for all effects (including the beam-induced effects reported in Ref.~\cite{Voutsinas:2019hwu}), are given in~\ref{sec:individualResults} for each of the four LEP experiments, with and without assuming lepton universality. With these updated parameters, and with the inclusion in Eq.~\ref{eq:Nnu} of the most accurate evaluation of $(\Gamma_{\nu\nu}/\Gamma_{\ell\ell})_{\rm SM}$ (predicted to amount to $1.99060 \pm 0.00021$ from the most up-to-date SM calculations of higher-order corrections to $\Gamma_{\nu\nu}$ and $\Gamma_{\ell\ell}$~\cite{DUBOVYK201886}, and the most recent measurements of the Higgs boson and top quark masses~\cite{Tanabashi:2018oca}), the number of light neutrino species measured by each LEP experiment is determined to be
\begin{eqnarray}
{\rm ALEPH} & : & N_\nu = 2.9994 \pm 0.0122, \\
{\rm DELPHI} & : & N_\nu = 2.9949 \pm 0.0163, \\
{\rm L3} & : & N_\nu = 2.9921 \pm 0.0133, \\ 
{\rm OPAL} & : & N_\nu = 2.9950 \pm 0.0128,
\end{eqnarray}
including a fully correlated uncertainty of $\pm 0.0032$ that comes from the common Bhabha cross-section theory error ($\pm 0.0028$), the uncertainty on the QED corrections to the Z lineshape ($\pm 0.0016$), and the uncertainty on $(\Gamma_{\nu\nu}/\Gamma_{\ell\ell})_{\rm SM}$ ($\pm 0.0003$). 

The combination of the four LEP experiments follows the exact same path, at each step, as that written up in Ref.~\cite{ALEPH:2005ab}. The combination code was checked to give the same result as in Ref.~\cite{ALEPH:2005ab} when starting from the original individual results, up to the last published digit, for all Z parameters, their uncertainties, and their correlations. The combined peak hadronic cross section and the corresponding number of light neutrino species obtained from Eq.~\ref{eq:NnuForLep} evolve as displayed in Table~\ref{tab:sighadnu} at each step. 

\begin{table}[htbp]
\centering
\caption{Combined peak hadronic cross section ($\sigma^0_{\rm had}$) and the corresponding number of light neutrino species $N_\nu$, at each step of the corrections considered in this letter.
\vspace{2mm}}
\label{tab:sighadnu}
\begin{tabular}{|l|c|c|}
\hline 
{\small Correction}          &  $\sigma^0_{\rm had}$ [nb]   & $N_\nu$                     \\\hline\hline
Original value  & $41.540\phantom{0} \pm 0.037\phantom{0}$ & $2.9846 \pm 0.0082$ \\ 
New  $(\Gamma_{\nu\nu}/\Gamma_{\ell\ell})_{\rm SM}$  & $41.5400 \pm 0.0372$ & $2.9856 \pm 0.0081$ \\ 
Z exchange          & $41.5390 \pm 0.0369$ & $2.9857 \pm 0.0080$\\
Light fermion-pairs & $41.5292 \pm 0.0353$ & $2.9875 \pm 0.0078$\\
Vacuum polarization & $41.5196 \pm 0.0324$ & $2.9893 \pm 0.0074$\\ 
Beam-induced        & $41.4802 \pm 0.0325$ & $2.9963 \pm 0.0074$\\ \hline
\end{tabular}
\end{table}

The updated combined Z parameters and correlations are given in~\ref{sec:combinationResults}, both with and without assuming lepton universality. Some excursion of the correlations between the peak hadronic cross section and the other Z parameters is observed. The other visible change is the slight increase of the Z width from 2.4952 to 2.4955\,GeV, as already noticed in Ref.~\cite{Voutsinas:2019hwu}. Finally, it is interesting to note that the increase of $N_\nu$ due to beam-induced effects is smaller than reported in Ref.~\cite{Voutsinas:2019hwu} ($\delta N_\nu = +0.0070$ instead of $+0.0072$). This is because the four LEP experiments are now affected by the same theoretical uncertainty, thus reducing OPAL's weight in the LEP combination. 


Extensive systematic studies are presented in Ref.~\cite{Voutsinas:2019hwu} for what concern the beam-induced effects. The corresponding uncertainties are propagated in this letter. Among those, the relative luminosity-increase uncertainties due to the LumiCal acceptance knowledge ($\pm 0.2\%$) and the averaging procedure over the whole LEP\,1 period ($\pm 0.5\%$), are common to all contributions, but not correlated among the LEP experiments. The (mostly statistical) uncertainties on the Z-exchange contribution and the light fermion-pair correction are indicated in Tables~\ref{tab:zrescale} and~\ref{tab:SoftPairCorrection}. The latter also contains systematic effects arising from the imperfect simulation of the detector granularity and the clustering procedure (which do not affect the Z-exchange and the vacuum polarization contributions). The inclusion of the 1990 data with nominal errors (instead of artificially inflating them by a large factor) changes $N_\nu$ by less than $+0.00001$ with the nominal correction, and by $-0.00004$ if no correction is applied for that year. These changes are not visible with the numbers of digits used in Table~\ref{tab:sighadnu}. The use of the DHMZ (KNT) hadronic vacuum polarization code instead of that of Jegerlehner would marginally change $N_\nu$ from $2.9963 \pm 0.00074$ to $2.9958 \pm 0.00074$ ($2.9954 \pm 0.00076$). Altogether, the above effects are responsible for an uncertainty on $N_\nu$ smaller than $\pm 0.0005$, already included in the estimate of Table~\ref{tab:sighadnu}. 


\section{Conclusions}
\label{sec:conclusion}

The modification of the Bhabha cross section and its uncertainty offered by recent theoretical developments has been quantified for the four experiments operating at LEP at and around the Z pole. The  integrated luminosity at the peak has been found to be underestimated by about $0.048\%$, a bias compatible with the theoretical uncertainty of $\pm 0.061\%$ reported at the time of LEP.  When this bias is corrected for -- on top of the beam-beam effect correction reported in Ref.~\cite{Voutsinas:2019hwu} -- the number of light neutrino species determined by the combined LEP experiments from the invisible decay width of the ${\rm Z}$ boson is determined to be  

\begin{equation*}
    N_\nu = 2.9963 \pm 0.0074, 
\end{equation*}
instead of the PDG value of $2.9840 \pm 0.0082$~\cite{ALEPH:2005ab,Tanabashi:2018oca}. The hadronic cross section at the Z peak 
and the Z width are also modified and become:
\begin{eqnarray*}
\sigma_{\rm had}^0 & = & 41.4802 \pm 0.0325\,{\rm nb},\\
\Gamma_{\rm Z} & = & \phantom{4}2.4955 \pm 0.0023\,{\rm GeV}.
\end{eqnarray*}
Correlations between the peak hadronic cross section and the other Z parameters are marginally modified, but no other electroweak precision observable is significantly affected. 

\newpage
\subsection*{Acknowledgements}
We are grateful to Maciej Skrzypek for having shared with us his irreplaceable expertise on the four-fermion real and virtual contributions, to Janusz Gluza for his tremendous insight and detailed code with the complete ${\cal O}(\alpha^2)$ corrections to the Bhabha cross section, and to Wies\l{}aw P\l{}aczek for useful discussions on the Z contribution. We express our appreciation to the DHMZ team (Michel Davier, Andreas H\"ocker, Bogdan Malaescu, Zhiqing Zhang) and the KNT team (Alexander Keshavarzi, Daisuke Nomura, Thomas Teubner) for providing us with the latest version of their private vacuum-polarisation code.  We are indebted to Gerardo Ganis for setting up working frameworks for the 15-to-30-years-old {\tt FORTRAN} codes used for this work, and to Emmanuel Perez for his critical reading of and refined comments to the manuscript. We thank Helmut Burkhardt, Martin Gr\"unewald, Michelangelo Mangano, Fulvio Piccinini, and Bolek Pietrzyk for interesting suggestions. This work is partly supported by the Polish National Science Centre grant 2016/23/B/ST2/03927.

\newpage
\appendix
\section{\small Fit results for ALEPH, DELPHI, L3, and OPAL}
\label{sec:individualResults}
\subsection{Assuming lepton universality}

{\setlength{\tabcolsep}{2.5pt} 
\renewcommand{\arraystretch}{1.1} 
\begin{table}[!htbp]
\caption{Individual experiment results on Z parameters and their correlation coefficients, 
assuming lepton universality. All systematic errors are included.
\vspace{2mm}}
\label{tab:LEPIndividualsLFU}
\begin{tabular}{|lrcr||rrrrr|}
\hline
\multicolumn{4}{|c||}{Parameters} & \multicolumn{5}{|c|}{Correlations} \\
\multicolumn{4}{|c||}{} & $m_{\rm Z}$ & $\Gamma_{\rm Z}$ & $\sigma^0_{\rm had}$ & $R^0_{\ell}$ & $A_{\rm FB}^{0,\ell}$ \\ \hline\hline
\multicolumn{9}{|c|}{ ALEPH } \\ \hline
$m_{\rm Z}$ [GeV] &  91.1893 & $\pm$ & 0.0031 & $\phantom{-}1.0000$ & & & & \\
$\Gamma_{\rm Z}$ [GeV] &   2.4962 & $\pm$ & 0.0043 & $0.0375$ & $\phantom{-}1.0000$ & & & \\
$\sigma^0_{\rm had}$ [nb] &  41.4919 & $\pm$ & 0.0539 & $-0.0905$ & $-0.3789$ & $\phantom{-}1.0000$ & & \\
$R^0_{\ell}$ &  20.7285 & $\pm$ &  0.0388 & $0.0314$ & $0.0110$ & $0.2430$ & $\phantom{-}1.0000$ & \\
$A_{\rm FB}^{0,{\ell}}$ &   0.0173 & $\pm$ & 0.0016 & $0.0724$ & $0.0020$ & $0.0023$ & $-0.0729$ & $\phantom{-}1.0000$ \\ \hline \hline
\multicolumn{9}{|c|}{ DELPHI } \\ \hline
$m_{\rm Z}$ [GeV] &  91.1862 & $\pm$ & 0.0028 & $\phantom{-}1.0000$ & & & & \\
$\Gamma_{\rm Z}$ [GeV] &   2.4878 & $\pm$ & 0.0041 & $0.0472$ & $\phantom{-}1.0000$ & & & \\
$\sigma^0_{\rm had}$ [nb] &  41.5201 & $\pm$ & 0.0665 & $-0.0695$ & $-0.2680$ & $\phantom{-}1.0000$ & & \\
$R^0_{\ell}$ &  20.7278 & $\pm$ &  0.0600 & $0.0276$ & $-0.0057$ & $0.2399$ & $\phantom{-}1.0000$ & \\
$A_{\rm FB}^{0,{\ell}}$ &   0.0186 & $\pm$ & 0.0019 & $0.0946$ & $0.0066$ & $-0.0050$ & $-0.0006$ & $\phantom{-}1.0000$ \\ \hline \hline
\multicolumn{9}{|c|}{ L3 } \\ \hline
$m_{\rm Z}$ [GeV] &  91.1894 & $\pm$ & 0.0030 & $\phantom{-}1.0000$ & & & & \\
$\Gamma_{\rm Z}$ [GeV] &   2.5031 & $\pm$ & 0.0041 & $0.0691$ & $\phantom{-}1.0000$ & & & \\
$\sigma^0_{\rm had}$ [nb] &  41.4690 & $\pm$ & 0.0508 & $0.0082$ & $-0.3392$ & $\phantom{-}1.0000$ & & \\
$R^0_{\ell}$ &  20.8079 & $\pm$ &  0.0598 & $0.0645$ & $0.0033$ & $0.1150$ & $\phantom{-}1.0000$ & \\
$A_{\rm FB}^{0,{\ell}}$ &   0.0193 & $\pm$ & 0.0025 & $0.0399$ & $0.0280$ & $0.0025$ & $-0.0249$ & $\phantom{-}1.0000$ \\ \hline \hline
\multicolumn{9}{|c|}{ OPAL } \\ \hline
$m_{\rm Z}$ [GeV] &  91.1853 & $\pm$ & 0.0030 & $\phantom{-}1.0000$ & & & & \\
$\Gamma_{\rm Z}$ [GeV] &   2.4952 & $\pm$ & 0.0041 & $0.0505$ & $\phantom{-}1.0000$ & & & \\
$\sigma^0_{\rm had}$ [nb] &  41.4462 & $\pm$ & 0.0518 & $0.0295$ & $-0.3481$ & $\phantom{-}1.0000$ & & \\
$R^0_{\ell}$ &  20.8214 & $\pm$ &  0.0447 & $0.0430$ & $0.0244$ & $0.2846$ & $\phantom{-}1.0000$ & \\
$A_{\rm FB}^{0,{\ell}}$ &   0.0145 & $\pm$ & 0.0017 & $0.0734$ & $-0.0052$ & $0.0140$ & $-0.0171$ & $\phantom{-}1.0000$ \\ \hline \hline

\end{tabular}
\end{table}
}

\subsection{Without assuming lepton universality}

{\setlength{\tabcolsep}{2.2pt} 
\renewcommand{\arraystretch}{1.1} 
\begin{table}[!htbp]
\fontsize{8.5}{10.2}\selectfont
\caption{Individual experiment results on Z parameters and their correlation coefficients,
without assuming lepton universality. All systematic errors are included.
\vspace{2mm}}
\label{tab:LEPIndividuals}
\begin{tabular}{|lrcr||rrrrrrrrr|}
\hline
\multicolumn{4}{|c||}{Parameters} & \multicolumn{9}{|c|}{Correlations} \\
\multicolumn{4}{|c||}{} & $m_{\rm Z}$ & $\Gamma_{\rm Z}$ & $\sigma^0_{\rm had}$ & $R^0_{\rm e}$ & $R^0_{\mu}$ & $R^0_{\tau}$ & $A_{\rm FB}^{0,{\rm e}}$ & $A_{\rm FB}^{0,\mu}$ & $A_{\rm FB}^{0,\tau}$ \\ \hline\hline
\multicolumn{13}{|c|}{ ALEPH } \\ \hline
$m_{\rm Z}$ [GeV] &  91.1891 & $\pm$ & 0.0031 & $\phantom{-}1.000$ & & & & & & & & \\
$\Gamma_{\rm Z}$ [GeV] &   2.4962 & $\pm$ & 0.0043 & $0.038$ & $\phantom{-}1.000$ & & & & & & & \\
$\sigma^0_{\rm had}$ [nb] &  41.4915 & $\pm$ & 0.0539 & $-0.090$ & $-0.379$ & $\phantom{-}1.000$ & & & & & & \\
$R^0_{\rm e}$ &  20.6900 & $\pm$ &  0.0751 & $0.101$ & $0.004$ & $0.133$ & $\phantom{-}1.000$ & & & & & \\
$R^0_{\mu}$ &  20.8010 & $\pm$ & 0.0561 & $-0.003$ & $0.012$ & $0.165$ & $0.086$ & $\phantom{-}1.000$ & & & & \\
$R^0_{\tau}$ &  20.7080 & $\pm$ & 0.0621 & $-0.003$ & $0.004$ & $0.150$ & $0.070$ & $0.097$ & $\phantom{-}1.000$ & & & \\
$A_{\rm FB}^{0,{\rm e}}$ &   0.0184 & $\pm$ & 0.0034 & $-0.047$ & $0.000$ & $-0.003$ & $-0.387$ & $0.000$ & $0.000$ & $\phantom{-}1.000$ & & \\
$A_{\rm FB}^{0,\mu}$ &   0.0172 & $\pm$ & 0.0024 & $0.072$ & $0.002$ & $0.002$ & $0.019$ & $0.013$ & $-0.000$ & $-0.007$ & $\phantom{-}1.000$ & \\
$A_{\rm FB}^{0,\tau}$ &   0.0170 &  $\pm$ & 0.0028 & $0.061$ & $0.002$ & $0.002$ & $0.017$ & $-0.000$ & $0.011$ & $-0.006$ & $0.017$ & $\phantom{-}1.000$ \\ \hline \hline
\multicolumn{13}{|c|}{ DELPHI } \\ \hline
$m_{\rm Z}$ [GeV] &  91.1864 & $\pm$ & 0.0028 & $\phantom{-}1.000$ & & & & & & & & \\
$\Gamma_{\rm Z}$ [GeV] &   2.4878 & $\pm$ & 0.0041 & $0.047$ & $\phantom{-}1.000$ & & & & & & & \\
$\sigma^0_{\rm had}$ [nb] &  41.5200 & $\pm$ & 0.0665 & $-0.069$ & $-0.268$ & $\phantom{-}1.000$ & & & & & & \\
$R^0_{\rm e}$ &  20.8800 & $\pm$ &  0.1201 & $0.063$ & $-0.000$ & $0.119$ & $\phantom{-}1.000$ & & & & & \\
$R^0_{\mu}$ &  20.6500 & $\pm$ & 0.0761 & $-0.003$ & $-0.007$ & $0.189$ & $0.056$ & $\phantom{-}1.000$ & & & & \\
$R^0_{\tau}$ &  20.8400 & $\pm$ & 0.1301 & $0.001$ & $-0.001$ & $0.112$ & $0.034$ & $0.053$ & $\phantom{-}1.000$ & & & \\
$A_{\rm FB}^{0,{\rm e}}$ &   0.0171 & $\pm$ & 0.0049 & $0.057$ & $0.001$ & $-0.006$ & $-0.106$ & $0.000$ & $-0.001$ & $\phantom{-}1.000$ & & \\
$A_{\rm FB}^{0,\mu}$ &   0.0165 & $\pm$ & 0.0025 & $0.064$ & $0.006$ & $-0.002$ & $0.025$ & $0.008$ & $0.000$ & $-0.015$ & $\phantom{-}1.000$ & \\
$A_{\rm FB}^{0,\tau}$ &   0.0241 &  $\pm$ & 0.0037 & $0.043$ & $0.003$ & $-0.002$ & $0.015$ & $0.000$ & $0.012$ & $-0.014$ & $0.015$ & $\phantom{-}1.000$ \\ \hline \hline
\multicolumn{13}{|c|}{ L3 } \\ \hline
$m_{\rm Z}$ [GeV] &  91.1897 & $\pm$ & 0.0030 & $\phantom{-}1.000$ & & & & & & & & \\
$\Gamma_{\rm Z}$ [GeV] &   2.5028 & $\pm$ & 0.0041 & $0.065$ & $\phantom{-}1.000$ & & & & & & & \\
$\sigma^0_{\rm had}$ [nb] &  41.4698 & $\pm$ & 0.0508 & $0.009$ & $-0.339$ & $\phantom{-}1.000$ & & & & & & \\
$R^0_{\rm e}$ &  20.8150 & $\pm$ &  0.0891 & $0.107$ & $-0.007$ & $0.074$ & $\phantom{-}1.000$ & & & & & \\
$R^0_{\mu}$ &  20.8610 & $\pm$ & 0.0971 & $-0.001$ & $0.002$ & $0.076$ & $0.032$ & $\phantom{-}1.000$ & & & & \\
$R^0_{\tau}$ &  20.7900 & $\pm$ & 0.1301 & $0.002$ & $0.005$ & $0.052$ & $0.025$ & $0.021$ & $\phantom{-}1.000$ & & & \\
$A_{\rm FB}^{0,{\rm e}}$ &   0.0107 & $\pm$ & 0.0058 & $-0.045$ & $0.055$ & $-0.006$ & $-0.146$ & $-0.001$ & $-0.003$ & $\phantom{-}1.000$ & & \\
$A_{\rm FB}^{0,\mu}$ &   0.0188 & $\pm$ & 0.0033 & $0.052$ & $0.004$ & $0.005$ & $0.017$ & $0.005$ & $0.000$ & $0.012$ & $\phantom{-}1.000$ & \\
$A_{\rm FB}^{0,\tau}$ &   0.0260 &  $\pm$ & 0.0047 & $0.034$ & $0.004$ & $0.003$ & $0.012$ & $0.000$ & $0.007$ & $-0.008$ & $0.007$ & $\phantom{-}1.000$ \\ \hline \hline
\multicolumn{13}{|c|}{ OPAL } \\ \hline
$m_{\rm Z}$ [GeV] &  91.1858 & $\pm$ & 0.0030 & $\phantom{-}1.000$ & & & & & & & & \\
$\Gamma_{\rm Z}$ [GeV] &   2.4952 & $\pm$ & 0.0041 & $0.049$ & $\phantom{-}1.000$ & & & & & & & \\
$\sigma^0_{\rm had}$ [nb] &  41.4460 & $\pm$ & 0.0518 & $0.031$ & $-0.348$ & $\phantom{-}1.000$ & & & & & & \\
$R^0_{\rm e}$ &  20.9010 & $\pm$ &  0.0841 & $0.107$ & $0.011$ & $0.153$ & $\phantom{-}1.000$ & & & & & \\
$R^0_{\mu}$ &  20.8110 & $\pm$ & 0.0581 & $0.001$ & $0.020$ & $0.219$ & $0.096$ & $\phantom{-}1.000$ & & & & \\
$R^0_{\tau}$ &  20.8320 & $\pm$ & 0.0911 & $0.001$ & $0.013$ & $0.135$ & $0.041$ & $0.054$ & $\phantom{-}1.000$ & & & \\
$A_{\rm FB}^{0,{\rm e}}$ &   0.0089 & $\pm$ & 0.0045 & $-0.053$ & $-0.005$ & $0.011$ & $-0.222$ & $-0.001$ & $0.005$ & $\phantom{-}1.000$ & & \\
$A_{\rm FB}^{0,\mu}$ &   0.0159 & $\pm$ & 0.0023 & $0.077$ & $-0.002$ & $0.011$ & $0.031$ & $0.018$ & $0.004$ & $-0.011$ & $\phantom{-}1.000$ & \\
$A_{\rm FB}^{0,\tau}$ &   0.0145 &  $\pm$ & 0.0030 & $0.059$ & $-0.003$ & $0.003$ & $0.015$ & $-0.010$ & $0.007$ & $-0.009$ & $0.014$ & $\phantom{-}1.000$ \\ \hline \hline

\end{tabular}
\end{table}
}
\newpage

\section{\small Fit results for the LEP combination}
\label{sec:combinationResults}
\subsection{Assuming lepton universality}
{\setlength{\tabcolsep}{2.6pt} 
\renewcommand{\arraystretch}{1.2} 
\begin{table}[htbp]
\begin{center}
\caption{LEP combined results on Z parameters and their correlation coefficients, 
assuming lepton universality. All systematic errors are included.
\vspace{2mm}}
\label{tab:LEPCombinedLFU}
\begin{tabular}{|lrcr||rrrrr|}
\hline
\multicolumn{4}{|c||}{Parameters} & \multicolumn{5}{|c|}{Correlations} \\
\multicolumn{4}{|c||}{} & $m_{\rm Z}$ & $\Gamma_{\rm Z}$ & $\sigma^0_{\rm had}$ & $R^0_{\ell}$ & $A_{\rm FB}^{0,\ell}$ \\ \hline\hline
\multicolumn{9}{|c|}{ LEP Combination } \\ \hline
$m_{\rm Z}$ [GeV] &  91.1875 & $\pm$ & 0.0021 & $\phantom{-}1.0000$ & & & & \\
$\Gamma_{\rm Z}$ [GeV] &   2.4955 & $\pm$ & 0.0023 & $-0.0228$ & $\phantom{-}1.0000$ & & & \\
$\sigma^0_{\rm had}$ [nb] &  41.4802 & $\pm$ & 0.0325 & $-0.0521$ & $-0.3248$ & $\phantom{-}1.0000$ & & \\
$R^0_{\ell}$ &  20.7666 & $\pm$ &  0.0247 & $0.0332$ & $0.0037$ & $0.1960$ & $\phantom{-}1.0000$ & \\
$A_{\rm FB}^{0,{\ell}}$ &   0.0171 & $\pm$ & 0.0010 & $0.0549$ & $0.0033$ & $0.0069$ & $-0.0560$ & $\phantom{-}1.0000$ \\ \hline \hline
\end{tabular}
\end{center}
\end{table}
}
\subsection{Without assuming lepton universality}
{\setlength{\tabcolsep}{2.1pt} 
\renewcommand{\arraystretch}{1.2} 
\begin{sidewaystable}[htbp]
\fontsize{11.2}{13.44}\selectfont
\begin{center}
\caption{LEP combined results on Z parameters and their correlation coefficients,
without assuming lepton universality. All systematic errors are included.
\vspace{2mm}}
\label{tab:LEPCombined}
\begin{tabular}{|lrcr||rrrrrrrrr|}
\hline
\multicolumn{4}{|c||}{Parameters} & \multicolumn{9}{|c|}{Correlations} \\
\multicolumn{4}{|c||}{} & $m_{\rm Z}$ & $\Gamma_{\rm Z}$ & $\sigma^0_{\rm had}$ & $R^0_{\rm e}$ & $R^0_{\mu}$ & $R^0_{\tau}$ & $A_{\rm FB}^{0,{\rm e}}$ & $A_{\rm FB}^{0,\mu}$ & $A_{\rm FB}^{0,\tau}$ \\ \hline\hline
\multicolumn{13}{|c|}{ LEP Combination } \\ \hline
$m_{\rm Z}$ [GeV] &  91.1876 & $\pm$ & 0.0021 & $\phantom{-}1.0000$ & & & & & & & & \\
$\Gamma_{\rm Z}$ [GeV] &   2.4955 & $\pm$ & 0.0023 & $-0.0238$ & $\phantom{-}1.0000$ & & & & & & & \\
$\sigma^0_{\rm had}$ [nb] &  41.4807 & $\pm$ & 0.0325 & $-0.0507$ & $-0.3249$ & $\phantom{-}1.0000$ & & & & & & \\
$R^0_{\rm e}$ &  20.8038 & $\pm$ &  0.0497 & $0.0783$ & $-0.0110$ & $0.1138$ & $\phantom{-}1.0000$ & & & & & \\
$R^0_{\mu}$ &  20.7842 & $\pm$ & 0.0335 & $-0.0000$ & $0.0079$ & $0.1391$ & $0.0694$ & $\phantom{-}1.0000$ & & & & \\
$R^0_{\tau}$ &  20.7644 & $\pm$ & 0.0448 & $0.0019$ & $0.0059$ & $0.0987$ & $0.0464$ & $0.0696$ & $\phantom{-}1.0000$ & & & \\
$A_{\rm FB}^{0,{\rm e}}$ &   0.0145 & $\pm$ & 0.0025 & $-0.0139$ & $0.0071$ & $0.0015$ & $-0.3704$ & $0.0013$ & $0.0029$ & $\phantom{-}1.0000$ & & \\
$A_{\rm FB}^{0,\mu}$ &   0.0169 & $\pm$ & 0.0013 & $0.0459$ & $0.0020$ & $0.0035$ & $0.0197$ & $0.0121$ & $0.0012$ & $-0.0242$ & $\phantom{-}1.0000$ & \\
$A_{\rm FB}^{0,\tau}$ &   0.0188 &  $\pm$ & 0.0017 & $0.0346$ & $0.0013$ & $0.0018$ & $0.0132$ & $-0.0030$ & $0.0093$ & $-0.0202$ & $0.0464$ & $\phantom{-}1.0000$ \\ \hline \hline

\end{tabular}
\end{center}
\end{sidewaystable}
}
\newpage

\bibliographystyle{jhep}
\bibliography{biblio}

\end{document}